\RequirePackage{fix-cm}
\documentclass[twocolumn,epjc3]{svjour3}  
\smartqed  
\RequirePackage{graphicx}
\RequirePackage[separate-uncertainty=true,multi-part-units=single]{siunitx}
\RequirePackage{subfig}
\RequirePackage{multirow}
\RequirePackage[colorlinks,citecolor=black,urlcolor=black,linkcolor=black]{hyperref}
\journalname{Eur. Phys. J. C}
\begin{document}

\title{Performance of short and long bent crystals for the TWOCRYST experiment at the Large Hadron Collider 
}



\author{
    L.~Bandiera\thanksref{INFN-Fe} \and
    R.~Cai\thanksref{CERN} \and
    S.~Carsi\thanksref{INFN-Mib,UniComo} \and
    S.~Cesare\thanksref{e1,INFN-Mi,UniMi} \and
    K.A.~Dewhurst\thanksref{e2,CERN,DCF} \and
    M.~D’Andrea\thanksref{CERN,INFN-LNL} \and
    D.~De~Salvador\thanksref{INFN-LNL,UniPd} \and
    P.~Gandini\thanksref{INFN-Mi} \and
    V.~Guidi\thanksref{INFN-Fe,UniFe} \and
    P.~Hermes\thanksref{CERN} \and
    G.~Lezzani\thanksref{INFN-Mib,UniComo} \and
    L.~Malagutti\thanksref{INFN-Fe,UniFe} \and
    D.~Marangotto\thanksref{INFN-Mi,UniMi} \and
    C.~Maccani\thanksref{CERN,INFN-Pd,UniPd} \and
    A.~Mazzolari\thanksref{INFN-Fe,UniFe} \and
    A.~Merli\thanksref{INFN-Mi,UniMi,EPFL} \and
    D.~Mirarchi\thanksref{CERN} \and
    P.~Monti-Guarnieri\thanksref{UniTri,INFN-Tri} \and
    C.E.~Montanari\thanksref{CERN,UoM} \and
    R.~Negrello\thanksref{INFN-Fe,UniFe} \and
    N.~Neri\thanksref{INFN-Mi,UniMi} \and
    M.~Prest\thanksref{INFN-Mib,UniComo} \and
    S.~Redaelli\thanksref{CERN} \and
    M.~Romagnoni\thanksref{INFN-Fe,UniFe} \and
    A.~Selmi\thanksref{INFN-Mib,UniComo} \and
    G.~Tonani\thanksref{INFN-Mi,UniMi} \and
    E.~Vallazza\thanksref{INFN-Mib} \and
    D.~Veres\thanksref{CERN,UniFft} \and
    F.~Zangari\thanksref{INFN-Mi,UniMi} \and
    M.~Zielińska\thanksref{CERN,WUT}
}

\thankstext{e1}{e-mail: Sara.Cesare@unimi.it}
\thankstext{e2}{e-mail: Kay.Dewhurst@cern.ch}

%

\institute{%
    INFN, Sezione di Ferrara , Ferrara, Italy \label{INFN-Fe} \and
    European Organization for Nuclear Research, CERN, Switzerland \label{CERN} \and
    INFN, Sezione di Milano Bicocca, Milan, Italy \label{INFN-Mib} \and
    Università degli Studi dell’Insubria, Como, Italy \label{UniComo} \and
    INFN, Sezione di Milano, Milan, Italy \label{INFN-Mi} \and
    Università degli Studi di Milano, Milan, Italy \label{UniMi} \and
    INFN, Laboratori Nazionali di Legnaro, Padua, Italy \label{INFN-LNL} \and
    Università degli Studi di Padova, Padua, Italy \label{UniPd} \and
    Università degli Studi di Ferrara, Ferrara, Italy \label{UniFe} \and
    INFN, Sezione di Padova, Padua, Italy \label{INFN-Pd} \and 
    EPFL, Lausanne, Switzerland \label{EPFL} \and
    The Warsaw University of Technology, Warsaw, Poland \label{WUT} \and
    Università degli Studi di Trieste, Trieste, Italy \label{UniTri} \and
    INFN, Sezione di Trieste, Trieste, Italy \label{INFN-Tri} \and
    The University of Manchester, Manchester, UK \label{UoM} \and
    Goethe-Universität Frankfurt, Frankfurt am Main, Germany \label{UniFft}
    \and
    \emph{Now at}  The Dalton Cumbrian Facility, The University of Manchester, Cumbria, UK \label{DCF} 
    \and
    \emph{Now also at} EPFL, Lausanne, Switzerland \label{EPFL}
}

\date{Received: date / Accepted: date}

\maketitle

\begin{abstract}
This study investigates the performance of bent silicon crystals intended to channel hadrons in a fixed-target experiment at the Large Hadron Collider (LHC). The phenomenon of planar channelling in bent crystals enables extremely high effective bending fields for positively charged hadrons within compact volumes. Particles trapped in the potential well of high-purity, ordered atomic lattices follow the mechanical curvature of the crystal, resulting in macroscopic deflections. Although the bend angle remains constant across different momenta (i.e., the phenomenon is non-dispersive), the channelling acceptance and efficiency still depend on the particle momentum.

Crystals with lengths in the range of \SI{5}{\centi\metre} to \SI{10}{\centi\metre}, bent to angles between \SI{5}{\milli\radian} and \SI{15}{\milli\radian}, are under consideration for measurements of the electric and magnetic dipole moments of short-lived charmed baryons, such as the $\Lambda_c^+$. Such large deflection angles over short distances cannot be achieved using conventional magnets.

The principle of inducing spin precession through bent crystals for magnetic dipole moment measurements was first demonstrated experimentally in the 1990s. Building on this concept, experimental layouts are now being explored for implementation at the LHC. The feasibility of such measurements depends, among other factors, on the availability of crystals that exhibit the required mechanical properties to reach the necessary channeling performance. To address this, a dedicated machine experiment—TWOCRYST—has been installed in the LHC to carry out beam tests in the \si{\tera\electronvolt} energy range. The bent crystals for TWOCRYST were fabricated and tested using both X-ray diffraction and high-momentum hadron beams at \SI{180}{\giga\electronvolt/c} at the CERN Super Proton Synchrotron (SPS) extraction lines. This paper presents an analysis of the performance of these newly developed crystals, as characterised by these measurements.

\keywords{bent crystal \and TWOCRYST \and LHC \and torsion \and channeling efficiency }
\end{abstract}

\section{Introduction}
\label{sec:intro}
Bent crystals can deflect high-energy particles through the planar channelling process~\cite{biryukov_crystal_1997}. Deflection by bent silicon crystals has been successfully utilised as part of the collimation system in the Large Hadron Collider (LHC) at CERN~\cite{bruning_lhc_2004,dandrea_characterization_2023}. Promising results have also been achieved recently for other applications, such as slow extraction~\cite{garattini_crystal_2022,velotti_septum_2019}, building on extensive experience with bent crystals in accelerators worldwide; see for example~\cite{murphy_first_1996,mazzolari_steering_2014,afonin_study_2016}. As part of CERN's Physics Beyond Colliders (PBC) initiative~\cite{jaeckel_quest_2020}, various applications of bent crystals for fixed-target (FT) implementations with multi-\SI{}{\tera\electronvolt} beams at the LHC have been studied~\cite{pbc-report2019}, including their use for beam extraction~\cite{afonin_schemes_2005} and integration into detection systems~\cite{mirarchi_layouts_2020}. 

One PBC FT study that generated particular interest aims to probe the spin of the charm quark. In this scheme, charmed baryons (e.g. $\Lambda_c^+$) are produced in a fixed target and subsequently channeled through a long bent crystal, inducing spin precession. The resulting spin rotation within the crystal’s strong effective electromagnetic fields enables measurements of the baryons’ electric and magnetic dipole moments~\cite{baryshevsky_possibility_2016,stocchi:523655,botella_search_2017,bagli_electromagnetic_2017,fomin_feasibility_2017,fomin_prospect_2020,aiola_progress_2021}. The experimental setup for this study utilises two bent crystals: a first, short, crystal to split protons from the LHC beam halo further away from the beam core and direct them onto the FT with sufficient separation from the main beam, and a second, long, crystal to channel the charmed baryons produced in the beam-target collisions. This double-crystal configuration is designed to enable FT collisions concurrently with the high-intensity proton operations of the LHC, without perturbing the proton-proton luminosity production at the main LHC experiments. A spin precession experiment based on this scheme, named ALADDIN (An LHC Apparatus for Direct Dipole moments INvestigation), is currently being studied for future deployment at the LHC~\cite{ALADDIN:LOI:2024}. 

\begin{figure}[htb]
\centering
\includegraphics[width=.48\textwidth]{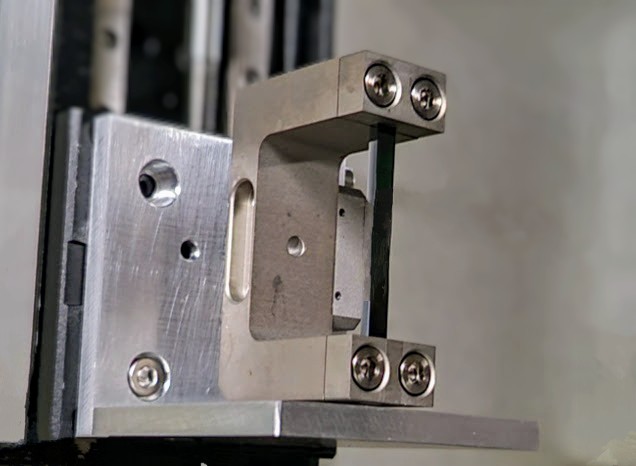}\\
\caption{TCCS bent silicon crystal clamped by its U-shaped metallic holder.}
\label{fig:tccs_crystal}       
\end{figure}

Channelling particles through two crystals in series (double-channelling) is a challenge requiring the precise positional and angular alignment of each crystal. Double-channelling has been demonstrated with \SI{450}{\giga\electronvolt} protons at CERN's Super Proton Synchrotron (SPS)~\cite{scandale_double-crystal_2021}. To prepare for spin precession experiments like ALADDIN, the proof-of-principle setup TWOCRYST~\cite{Hermes:ichep24:twocryst} has been implemented in the LHC to make the first demonstration of proton double-channelling at~\SI{}{\tera\electronvolt} energies in 2025. TWOCRYST incorporates two bent crystals with specifications optimised for a full-fledged experiment. The first crystal, the TCCS (Target Collimator Crystal for Splitting), has the same parameters as existing LHC collimator crystals, with a length of \SI{4}{\milli\meter} and a bend angle of approximately \SI{50}{\micro\radian}~\cite{mirarchi_crystal_2015-1}. A photograph of a TCCS crystal is shown in Fig.~\ref{fig:tccs_crystal}. The second crystal, the TCCP (Target Collimator Crystal for Precession), is significantly longer (\SI{70}{\milli\meter}), with a bend angle over $100$ times larger (\SI{6.9}{\milli\radian}), making it suitable for studying the precession of short-lived baryons. 

TWOCRYST provides the first opportunity to characterise such long crystals in the \SI{}{\tera\electronvolt} energy range. Particularly the TCCP, with its length and bend angle being unprecedented for the LHC, must be well characterised before installation into the collider. Pre-installation tests include detailed mechanical and visual inspections, X-ray characterisation, validation of mechanical stability against thermal cycles, mimicking the bake-out procedures required for integration into the ultra-high vacuum environment of the LHC, and assessments of channelling efficiency with hadron beams. This paper reports the results of the validation of three bent silicon crystals supplied by INFN-Ferrara for TWOCRYST, with emphasis on measurements using high-resolution X-ray diffraction (HR-XRD)~\cite{mazzolari_bent_2018} and hadron beams, carried out using the H8 beamline in CERN's North Area with a telescopic detector~\cite{lietti_microstrip_2013}. The analysis of hadron beams in H8 follows a well-established approach, see for example~\cite{garattini_overview_2018}. Other recent measurements with crystals producing bends in the multi-\SI{}{\milli\radian} range and the solutions adopted for the data analyses can be consulted in~\cite{aiola_progress_2021,rossi_track_2021}.

Three bent crystals were produced and tested for TWOCRYST: one for splitting and two for precession, each using different bending technologies. This paper describes the design of these crystals and presents the results of the characterisation measurements carried out before installation in the LHC. Both X-ray and particle beam measurements are used to determine the bend angle $\theta_{b}$ and its variation across the entrance face of each crystal (quantified by the torsion $\tau$). Particle beam measurements also provide the channelling efficiency $\epsilon_\text{ch}$ of the crystals, which can be compared with simulations. These results, particularly the channelling efficiency, are used to assess the suitability of each crystal for incorporation into the TWOCRYST experimental setup.

\section{Bent crystal designs}
The three crystals measured as part of the TWOCRYST effort include a baseline TCCS and TCCP, which are mounted using conventional metallic holders, and a third crystal using a novel anodic bonding technique developed at INFN-Ferrara. We refer to this third crystal as the TCCPA: Target Collimator Crystal for Precession - Anodic. Anodic bonding is expected to be beneficial for bending longer crystals, as it promises a more homogeneous bending curvature and improved torsion control.

\begin{figure}[htb]
\centering
\includegraphics[width=.48\textwidth]{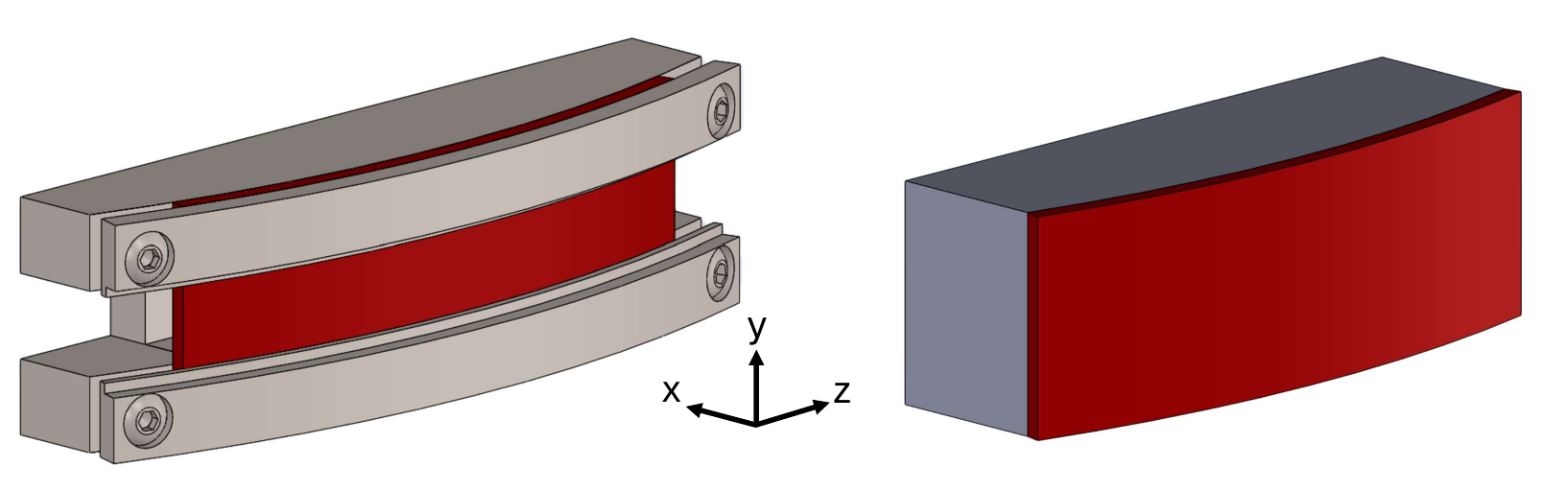}\\
\caption{Supports for the deformation of the silicon crystal wafers. The TCCP (left) is clamped in a specially-shaped metallic holder. The TCCPA (right) is anodically bonded to a silicon-based glass lens~\cite{lapadatu_chapter_2015}.}
\label{fig:crystal_holders}       
\end{figure}

All three crystals are manufactured from prime silicon wafers. The TCCS bending mechanism exploits the anticlastic deformation of a `strip' crystal~\cite{mazzolari_silicon_2021,baricordi_shaping_2008} while the TCCP and TCCPA crystals are designed to induce bending directly along the crystal surface. Both the baseline TCCS and TCCP are deformed by clamping them into specially shaped supports; a technique first used in experiments designed to extract the beam circulating in the SPS at CERN~\cite{akbari_first_1993}. These metallic holders with U-shaped cross-sections are shown in Figures~\ref{fig:tccs_crystal} (TCCS) and~\ref{fig:crystal_holders} (TCCP). The TCCP holder design was optimised to minimise aperture constraints for the LHC circulating beam, requiring special machining of the clamps facing the beam, see Figure~\ref{fig:crystal_holders} (left). For the TCCPA, the silicon wafer is bonded~\cite{lapadatu_chapter_2015} to the cylindrically curved surface of a silicon-based glass lens. The TCCPA, shown in Figure~\ref{fig:crystal_holders} (right), is a long crystal of approximately \SI{70}{\milli\meter} but has a smaller bending radius than the TCCP and, therefore, a stronger deflection of around~\SI{13}{\milli\radian}. This study reports the first measurement of an anodic bonded crystal using a high-energy hadron beam. The design properties of the three crystals are summarised in Table~\ref{tab:3crystals}.

\begin{table}[htb]
\caption{Design properties of the three crystals tested for TWOCRYST. The expected properties of the TCCS and TCCP are taken from the functional specification~\cite{demassieux_tccstccp_2022}.}
\resizebox{1.0\columnwidth}{!}{%
\label{tab:3crystals}
\begin{tabular}{lcccc}
\hline\noalign{\smallskip}
Crystal & & TCCS & TCCP & TCCPA  \\
\noalign{\smallskip}\hline\noalign{\smallskip}
Crystal material & & Si & Si & Si \\
Bending plane & & (110) & (110) & (110) \\
Length & {[}mm{]} & 4 & 70 & 70.5 \\
Width & {[}mm{]} &  35 & 8 & 22.5 \\
Height  & {[}mm{]} &  2 & 2 & 2 \\
Bend radius $\rho$ & {[}m{]} & 80 & 10 & 5.3 \\
Bend angle $\theta_b$ & {[}\SI{}{\milli\radian}{]} & 0.05 & 7.0 & 13.3 \\
$\theta_L$ at 180~\SI{}{\giga\electronvolt\per c} & {[}\SI{}{\micro\radian}{]}  & {13.3} & {12.9} & {12.5} \\
\noalign{\smallskip}\hline
\end{tabular}}
\end{table}



\section{Channelling in bent crystals}
\label{sec:channelling_theory}
Charged particles passing through a crystal with an ordered lattice structure experience an intense electromagnetic field, for example, of the order \SI{5}{\giga \electronvolt\per\centi\meter} in silicon~\cite{biryukov_crystal_1997}. As theoretically explained by Lindhard~\cite{lindhard_influence_1965}, when a charged particle has a small incident angle, $\theta_\text{in}$, relative to the crystallographic planes or axes, it experiences the effect of a continuous potential that traps the particle between the planes. These phenomena are called planar and axial channelling, respectively. 

For channelling to occur, the incident angle must be smaller than the Lindhard (or critical) angle, beyond which the transverse momentum exceeds the height of the potential well. In a straight crystal, the Lindhard angle is given by~\cite{biryukov_crystal_1997}
\begin{equation}\label{eq:lindhard}
    \theta_{L}=\sqrt{\frac{2 \, U_{0}}{p \, \beta \, c}} \,,
\end{equation} 
where $U_0$ is the potential well depth, $p$ is the particle momentum, $\beta$ is the relativistic particle velocity, and $c$ is the speed of light in vacuum. Positively charged particles can undergo planar channelling in crystals with a diamond cubic structure (e.g., C, Si, Ge) and will follow the naturally straight crystalline lattice~\cite{biryukov_crystal_1997}. By mechanically bending the crystal, the lattice planes become curved, causing any particles channelled between the planes to follow this curved trajectory. Therefore, it is possible to deflect particles by an angle equal to the crystal bending angle $\theta_b$.

For bent crystals, the effective potential-well depth decreases with increased bending per unit length~\cite{biryukov_crystal_1997}. The potential-well depth $U_{0}$ scales with a factor $(1-\rho_{c}/\rho)^2$, where $\rho$ is the bending radius and $\rho_{c}$ is the critical bending radius, beyond which channelling cannot occur. The Lindhard angle then becomes
\begin{eqnarray}
\theta_{L} &=&  \left(1-\frac{\rho_{c}}{\rho}\right)\sqrt{\frac{2U_{0}}{E}}\;\label{eq:critical_angle1},
\end{eqnarray}
with the critical radius being
\begin{eqnarray}
\rho_{c} &=& \frac{E}{U'(x_c)} 
\label{eq:critical_angle2}.
\end{eqnarray}
$U'(x_c)$ is given as \SI{5.7}{\giga\electronvolt\per\centi\meter} for the (110) plane in silicon~\cite{biryukov_crystal_1997}, using the Moli\`ere approximation for the potential~\cite{moliere_theorie_1947}. We assume $U_{0}=\SI{16}{\electronvolt}$ for silicon~\cite{biryukov_crystal_1997}. This results in Lindhard angles of \SI{13.3}{\micro\radian} for the TCCS, \SI{12.9}{\micro\radian} for the TCCP and \SI{12.5}{\micro\radian} for the TCCPA, at $E= p \, \beta \, c = \SI{180}{\giga\electronvolt \per c}$. These calculated values are also listed for reference in Table~\ref{tab:3crystals}.

\begin{figure}[htb]
\centering
\includegraphics[width=.47\textwidth]{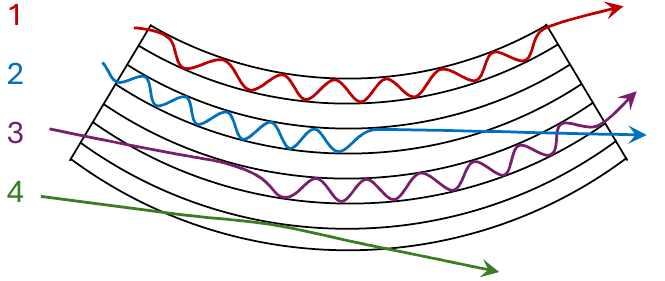}\\
\caption{Possible particle interactions with the lattice channels of a bent crystal. These are labelled 1) channelling, 2) dechannelling, 3) volume capture, and 4) volume reflection. A fifth interaction, amorphous (Coulomb) scattering also occurs independently of the channelling planes.}
\label{fig:interactions}       
\end{figure}

Even if impacting conditions for channelling are respected, particles may undergo other interactions, for example, amorphous (Coulomb) scattering, dechannelling, volume capture, or volume reflection~\cite{biryukov_crystal_1997}. Figure~\ref{fig:interactions} illustrates these different interactions. Amorphous scattering is an elastic interaction between an incoming particle and a lattice atom; a possible interaction process with any bulk material. Dechannelling occurs when an initially channelled particle escapes the potential well before reaching the end of the crystal; the particle trajectory is deflected by $<\theta_b$. Each crystal in this study has a length greater than the nuclear dechannelling length ($\sim\SI{1}{\milli\meter}$ at momenta of \SI{180}{\giga\electronvolt \per c}) and shorter than the electronic dechannelling length ($\sim\SI{100}{\milli\meter}$ at \SI{180}{\giga\electronvolt\per c}), as shown in Table~\ref{tab:3crystals}. In this regime, which is still not fully explored experimentally, dechannelling is predominately caused by multiple scattering with the atomic nuclei~\cite{scandale_observation_2009}. Volume capture allows a particle initially outside the channelling condition to enter channelling, e.g., by losing energy through scattering. Volume reflection can occur when a particle approaches the crystal bending plane close to the tangent, i.e., when the incident angle is in the range $\theta_L<\theta_\text{in}<\theta_b$. Volume reflection causes particles to be reflected in a direction opposite to the bending direction of the crystal.


In a final fixed-target experiment, achieving sufficient precession is essential for maximising statistics. This requires both the TCCS and TCCP crystals to have suitable channelling efficiencies. The properties of the TCCS crystals are well-established, thanks to experience gained from their use for LHC beam collimation~\cite{dandrea_characterization_2023}. However, the performance of the novel TCCP crystals presents an unprecedented challenge. In both cases, the channelling efficiency, $\epsilon_\text{ch}$, is defined as the fraction of particles channelled by the crystal, $N_\text{ch}$, relative to the total number of particles with the potential to be channelled, $N_\text{tot}$, meaning those with an incoming angular range within one Lindhard angle or $\pm\frac{1}{2}\theta_{L}$:

\begin{equation}\label{eq:efficiency}
    \epsilon_\text{ch} [\%]= \frac{N_\text{ch}}{N_\text{tot}}\times100 .
\end{equation}

\section{Characterising the crystals with X-rays}
\label{sec:xray}
Before the hadron beam tests, the deformation of the crystal wafers was measured using high-resolution X-ray diffraction (HR-XRD)~\cite{mazzolari_bent_2018}. This technique is used to assess the deformation of the front surface and the uniformity of the bending radius. Deformations of the entry surface cause variations in the optimal crystal-beam angular alignment as a function of the transverse impact positions. A preliminary X-ray assessment is crucial in selecting an appropriate crystal for the hadron tests, particularly for the long crystals. However, the entry surface deformation is best observable in the hadron beam tests, as shown in Figure~\ref{fig:torsion}.

\begin{figure}[htb]
\centering
\includegraphics[width=.47\textwidth]{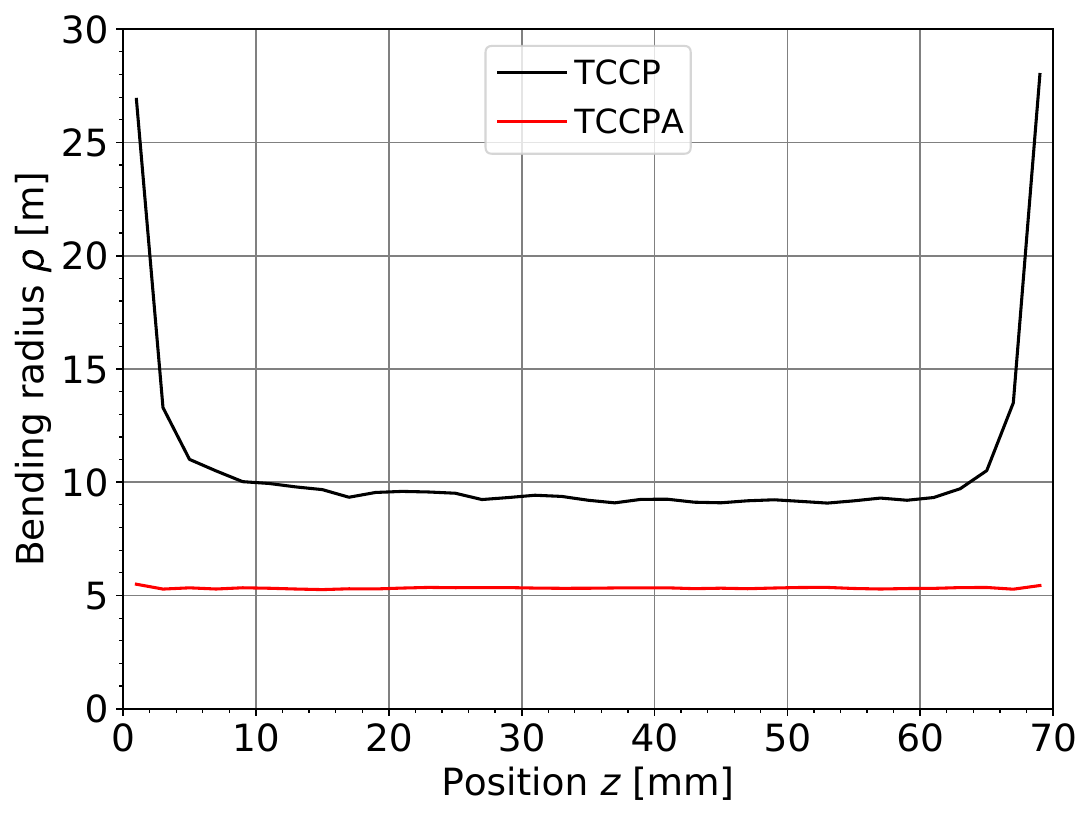}\\
\caption{Bending radius along the crystal in the direction of the beam, $z$. The measured bending radii are shown for both long crystals; TCCP (black) and TCCPA (red).}
\label{fig:bend_radius}
\end{figure}

Non-uniformities in the bending radius typically increase the risk of dechannelling of particles that respect channelling conditions at the entrance of the crystal~\cite{jensen_deflection_1992,gibson_deflection_1984}. The results of bending radius measurements using X-rays along the crystal length are shown for both TCCP and TCCPA in Figure~\ref{fig:bend_radius}. The uniformity of the bending radius along the full length of the TCCPA crystal is remarkable. In contrast, the TCCP bending radius varies greatly along the length of the crystal. At both crystal ends , the bending radius exceeds twice the target value of $R=\SI{10}{\meter}$. While these larger values are not a priori a direct cause of channelling efficiency loss, this aspect will be studied further.

\section{Characterising the crystals with 180~GeV hadron beams}
\label{sec:hadron_measurements}
\subsection{Experimental setup}
Each crystal was tested at the H8 beamline in the North Area at CERN, which is driven by \SI{400}{\giga\electronvolt} protons extracted from the SPS. This beamline operates with a beryllium target, and a spectrometer selecting \SI{180}{\giga\electronvolt} hadrons~\cite{al_measurement_1999}. These secondary beams are roughly composed of 30\% $\pi^+$ mesons and 70\% protons.

\begin{figure}
\includegraphics[width=.47\textwidth]{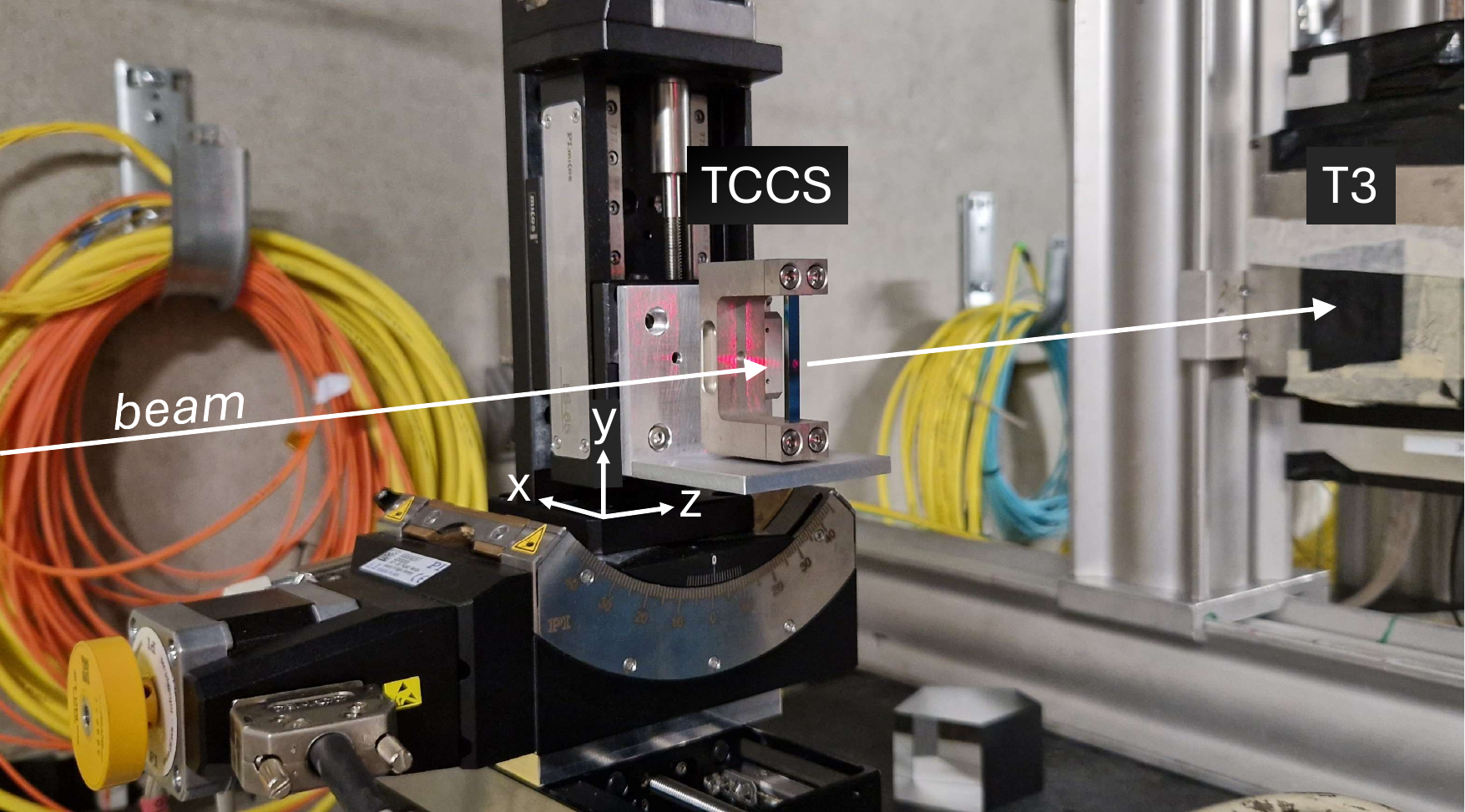}
\caption{Initial laser alignment of the TCCS crystal. The crystal is clamped into a bent position by the U-shaped metallic holder and mounted on a goniometer. The red laser light is visible on the crystal and holder. The silicon detector (T3) is visible downstream.}
\label{fig:mounted_crystal} 
\end{figure}

\begin{figure}
  \includegraphics[width=.47\textwidth]{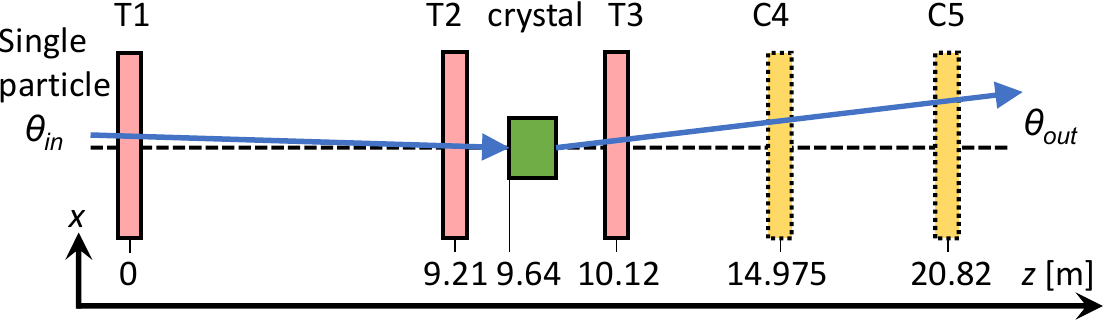}
\caption{Experimental layout employed on the H8 beamline for the crystal test. Two telescope detectors (T) are located upstream of the crystal and used to track the position and incidence angle $\theta_\text{in}$ of the incoming particles. A telescope (T) and a chamber (C) detector are located downstream of the crystal. Chamber C5 was used with the TCCS crystal, whereas chamber C4 was used with the larger bend-angle crystals; the TCCP and TCCPA. The trajectory of a single particle deflected by the crystal (blue) is shown passing through the detectors. The $z$-positions are given at the detectors' centres and the crystals' upstream edge.}
\label{fig:apparatus}
\end{figure}

For the characterization with hadrons, each crystal is mounted on a goniometer, shown in Figure~\ref{fig:mounted_crystal}, which provides precision control of the linear and angular crystal position relative to the incoming beam. An initial alignment of each crystal with the hadron beam is carried out using an alignment laser, mirror and beam splitter. Each crystal is mounted so that channelled particles are deflected in the $x$-plane. Figure~\ref{fig:mounted_crystal} shows the TCCS crystal during the laser alignment. Four silicon detectors, two upstream and two downstream of the crystal are used to track the incoming and outgoing particle trajectories. The distance between the two downstream detectors is adjusted to optimise measurements based on each crystal's bend angle. For crystals with larger bend angles, up to approximately \SI{14}{\milli\radian}, a shorter distance is used to ensure accurate tracking of the deflected particles. This method follows an established approach for measuring silicon crystals at H8~\cite{rossi_measurements_2015,hall_high_2019,scandale_study_2018}. Figure~\ref{fig:apparatus} shows an overview of the experimental layout.

Silicon microstrip detectors of two types were used: telescope detectors (T) and beam chamber detectors (C). The telescope detectors comprise \SI{384}{} strips, with a \SI{50}{\micro\meter} strip width, providing a \SI{1.92}{\centi\meter\squared} sensitive area and \SIrange{5}{10}{\micro\meter} resolution, depending on orientation~\cite{lietti_microstrip_2013,bonfanti_high_2012}. The beam chamber detectors comprise \SI{384}{} strips, with a~\SI{242}{\micro\meter} strip width, providing a sensitive area of~\SI{9.3}{\centi\meter\squared} and spatial resolution of~\SI{30}{\micro\meter}~\cite{barbiellini_agile_2002,prest_agile_2003}. Telescope detectors were used as the first three detectors (T1, T2, T3) to benefit from their fine spatial resolution. A chamber detector (C4 or C5) was used as the most downstream detector; exploiting the larger sensitive area to capture both the main beam and the channelled beam of hadrons. Chamber C4, located closer to the crystal, was used with the larger bend-angle crystals (TCCP and TCCPA) and C5, further downstream, was used with the TCCS crystal.

After an initial laser alignment, three sets of measurements were collected for each crystal, yielding the \textit{reference}, \textit{amorphous}, and \textit{channelling} datasets. To obtain the \textit{reference} dataset, the crystal was moved out of the beam, and data were recorded without the crystal present—allowing the beam profile to be observed directly in each detector. The crystal was then inserted, and a linear scan was performed to centre it within the beam. Once centred, the crystal was slightly rotated to induce amorphous scattering. Since only particles impacting the bulk of the crystal are scattered, the crystal edges could be identified from this \textit{amorphous} dataset. Subsequently, the optimal channelling orientation was determined via an angular scan. Finally, while the crystal was aligned in this optimal orientation, high-statistics \textit{channelling} data ($>$ \SI{1e6}{} events) were recorded. This procedure was repeated for each of the three crystals.


\subsection{Analysis procedure}
The collected data were analysed independently at CERN and INFN Milano, compared and validated. The data analysis follows a three-step approach. In the first step, the \textit{reference} data, collected without interactions with the crystal, are used to find the relative positions of the tracking detectors. This is done either by defining the origin (0,0) position of each tracker in ($x$,$y$) from the mean position of the passing particles, or using an alignment algorithm, based on Millepede-II~\cite{klienwort_millepede-ii_2024,blobel_software_2006}, which additionally takes into account the rotation of the three downstream detectors (T2, T3, C4/5) in the $x$-$y$ plane.

In the latter approach, the alignment parameters $\vec{\mu}$ are the \textit{x} and \textit{y} offsets of the last three detectors and their rotations around the \textit{z} axis. The coordinates at the first detector and the longitudinal \textit{z} positions of the other three detectors are fixed. The alignment parameters are determined with a $\chi^2$ minimisation,
\begin{equation}\label{eq:chi_squared}
    \chi^2 = \frac{|\vec{x}_\text{m}-\vec{x}_\text{e}(\vec{\mu})|^2}{\sigma_{\vec{x}}^2},
\end{equation} 
where $\vec{x}_\text{m}$ is the measured hit position for each track, $\vec{x}_\text{e}$ is the expected hit position based on the parameters $\vec{\mu}$, and $\sigma_{\vec{x}}$ is the experimental position resolution. 
The longitudinal \textit{z} translation of the detectors and their rotations around the \textit{x} and \textit{y} axes are second-order effects, therefore, they are not taken into consideration. The parameters $\vec{\mu}$ obtained by this approach are subsequently used to correct the detector positions for the analysis of the datasets recorded with crystals in place.

In a second step, the data recorded with crystals in \textit{amorphous} orientation are used to find the edge of each crystal in the deflection plane $x$. The incoming track of each particle passing T1 and T2 is tracked to the $z$-position of the upstream face of the crystal to give the incoming $x$-positions. Each particle's incoming and outgoing angles are similarly determined using position data from each pair of upstream and downstream detectors. Plotting the deflection $\Delta\theta_{x}$ against position $x$, shows a clear transition between weakly deflected particles (due to multiple scattering in the air or detector materials) and those with significant deflection due to amorphous scattering with the crystal. This process identifies the crystal edges in the $x$-plane and verifies the measured width (\SI{20}{\milli\meter}) of each crystal. All particles outside the defined crystal edges, and thus with zero channelling probability are removed from the \textit{channelling} dataset.

Finally, the high statistics \textit{channelling} data, collected with the crystal in the channelling orientation, are used to determine the channelling efficiency for the three crystals, as defined in Equation~\ref{eq:efficiency}. The parameter $N_\text{tot}$, only includes those particles with the potential to be channelled, i.e., those impacting the crystal with an incoming angle in the bending plane within one Lindhard angle; $\pm\frac{1}{2}\theta_{L}$. As the Lindhard angle $\theta_L$ depends on the crystal properties, it differs for each crystal. The values of $\theta_L$ for the three crystals relative to \SI{180}{\giga\electronvolt\per c} pions, and calculated following \cite{biryukov_crystal_1997}, are reported in Table~\ref{tab:3crystals}. These values are used to remove particles with unsuitable angles from the \textit{channelling} dataset and in turn the channelling efficiency calculations.

\begin{figure}[htb]
  \centering
  \includegraphics[width=.47\textwidth]{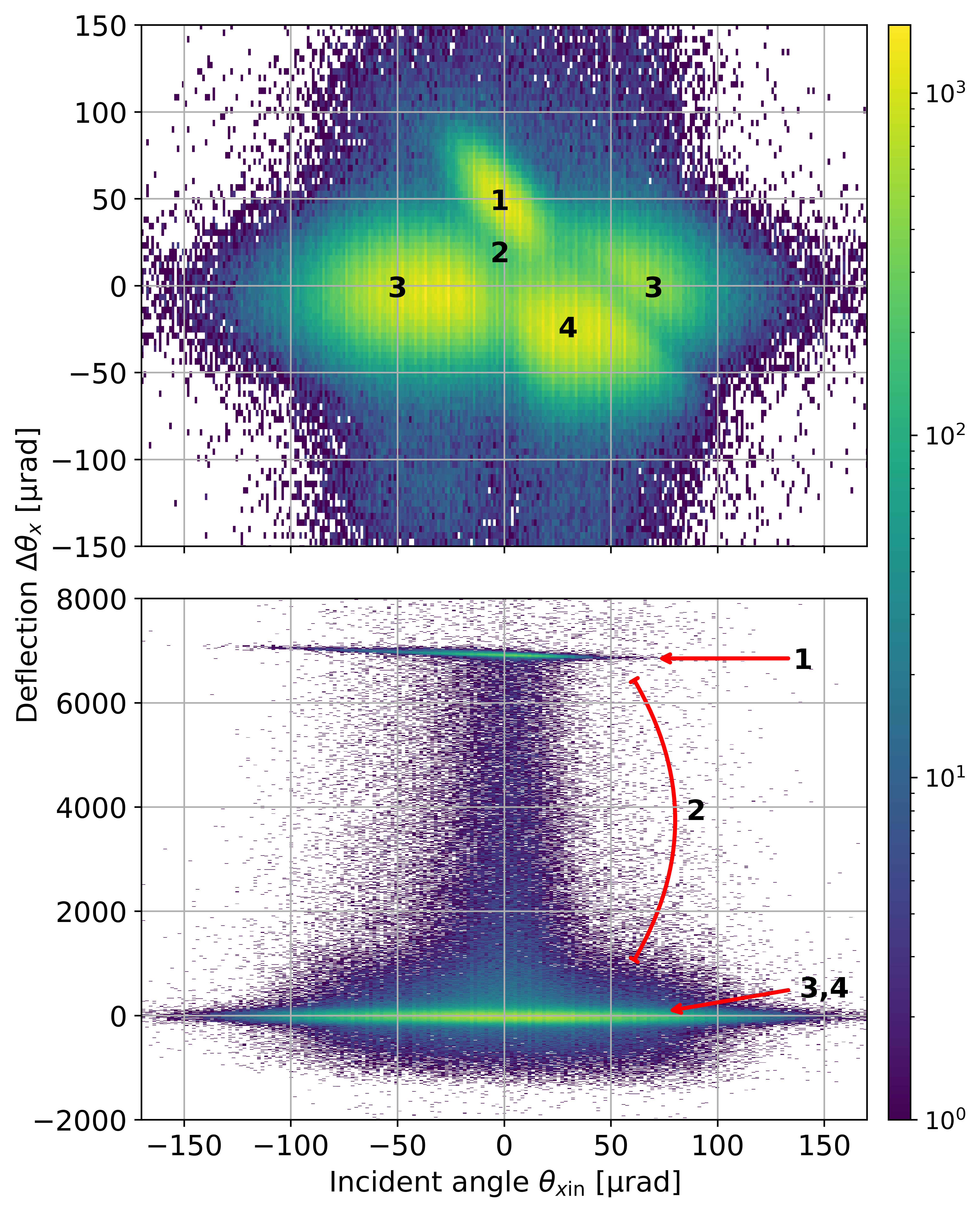}\\
\caption{Angular deflection of the particles impacting the crystals as a function of the incident angle to the crystalline plane. The upper plot shows the TCCS and the lower plot shows the TCCP. The colour scale shows the particle count in each pixel. The different interaction regimes are labelled 1) channelling, 2) dechannelling, 3) amorphous, and 4) volume reflection. The separation between amorphous and volume-reflected regions is less obvious in the lower plot due to the scale required for the large (\SI{}{\milli\radian}) deflection of the TCCP.}
\label{fig:channelling_regions}       
\end{figure}

The analysis method for the channelling efficiency varies among the three crystals due to their distinct characteristics. Specifically, the effect of torsion, $\tau_y$ -- the variation of the crystalline plane orientation along the direction ($y$) perpendicular to the bending plane ($x$) -- must be accounted for when measuring the channelling efficiency. The different orientations of the crystallographic planes along the crystal front surface influence the angular range in which particles can be channelled, i.e., the cut at $\pm\frac{1}{2}\theta_{L}$ may not be centred on zero, but a correction parameter $\theta_0=\theta_0(y)$. The TCCS crystal has the smallest bending angle, a smaller impact surface and negligible torsion by construction; values are typically smaller than \SI{1}{\micro\radian\per \milli\meter}. For this reason, the channelling efficiency can be extracted from the whole sample of particles impacting the crystal surface within $\theta_0\pm\frac{1}{2}\theta_{L}$, giving $N_\text{tot}$. The number of channelled particles $N_\text{ch}$ can be identified from their deflection angle $\Delta\theta_{x}$, as shown by the labelled regions in Figure~\ref{fig:channelling_regions}. The channelling efficiency is calculated in several iterations varying $\theta_0$ until the highest efficiency is found; representing the best alignment of incoming particles with the crystalline planes. For the longer crystals, the TCCP and TCCPA, the effect of the torsion is not negligible and must be incorporated into the analysis.

In the cases of the TCCP and TCCPA crystals, the effect of the torsion has been corrected by dividing the data sample into square bins, as a function of the impact position on the crystal surface. It is noted that in the case of the precession crystals, $\theta_0=\theta_0(x,y)$. The deflection angle $\Delta\theta_{x}$, is used to identify the channelled particles for each bin. The angular shift parameter $\theta_0$ is then varied to find the maximum channelling efficiency $\epsilon_\text{ch}$ for each bin; producing a 2D map of torsion variation across the crystal surface. The effect of torsion is then corrected using the parameter $\theta_0$ found for each bin. 

Examples of measured data are given in Figure~\ref{fig:channelling_regions} for the TCCS (top) and TCCP (bottom). The different particle interactions can be identified in overlapping regions on the plots of deflection $\Delta\theta_{x}$ against incoming angle $\theta_{x\text{in}}$.

\subsubsection{Splitting crystal (TCCS) results}
\begin{figure}[htb]
\centering
\includegraphics[width=.48\textwidth]{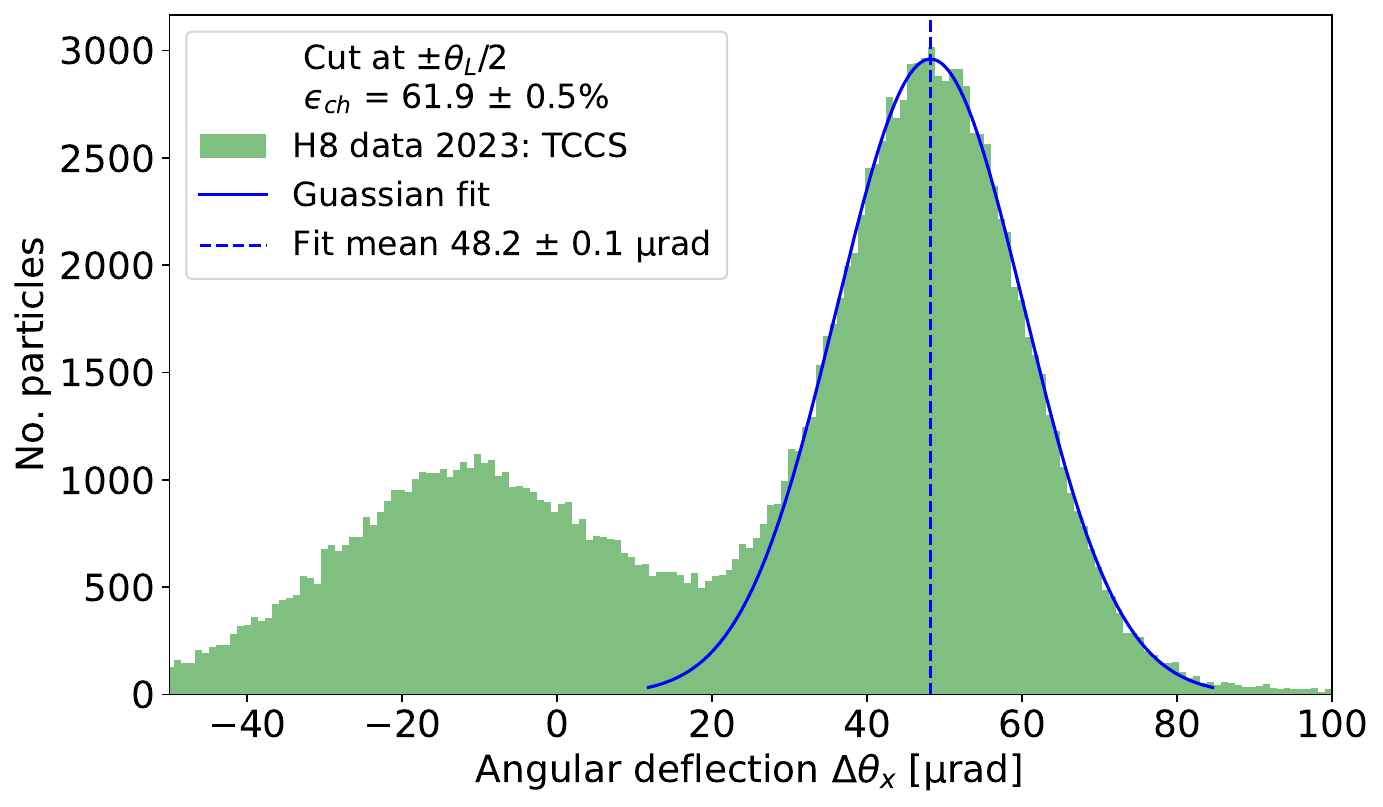}
\caption{Distribution of the angular deflection of particles which impact the TCCS crystal within half the Lindhard angle $(13.3/2)\SI{}{\micro\radian}$. The bin width is \SI{0.9}{\micro\radian}, calculated from $3.49\sigma/(N^{\frac{1}{3}})$~\cite{scott_optimal_1979,gholamy_what_2017}, where $\sigma$ is the standard deviation of the fit and $N$ is the sample size in the fitted peak. The right peak includes the channelled particles. Dechannelled particles receive a deflection between zero and $\theta_b$ so are spread between the two peaks. A Gaussian fit made to the right side of the channelled peak (from \SI{40}{\micro\radian} to \SI{80}{\micro\radian}) has a mean of \SI{48.2(0.1)}{\micro\radian} and standard deviation of \SI{12.1}{\micro\radian}. The area enclosed by the full Gaussian curve contains \SI{61.9(0.5)}{\%} of the impacting particles, revealing the channelling efficiency of the TCCS.}
\label{fig:TCCS_ch_eff}
\end{figure}

Figure~\ref{fig:TCCS_ch_eff} shows the deflection angle distribution for the TCCS crystal. The channelled particle peak is visible centred on \SI{48}{\micro\radian}, close to the design bending angle $\theta_b$ of \SI{50}{\micro\radian}~\cite{demassieux_tccstccp_2022}. The left peak contains a distribution of amorphously-scattered and volume-reflected particles. Dechannelled particles occupy the region between the two peaks. A Gaussian fit is made to the right side of the channelled peak, to reduce any contribution from dechannelled particles. The area under the fitted Gaussian gives the number of channelled particles $N_\text{ch}$ for the channelling efficiency calculation. The channelling efficiency of the TCCS, derived from the measurement using this approach, yields $\epsilon_\text{ch}^\text{TCCS} = 61.9 \pm 0.5 \%$. This value is also reported in Table \ref{tab:3crystals_meas}.

\subsubsection{Precession crystal (TCCP \& TCCPA) results}
\begin{figure}[!h]
    {\includegraphics[width=.47\textwidth]{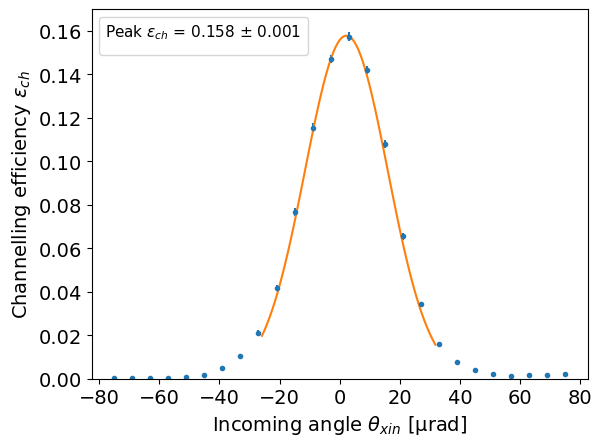}} \quad\quad
    {\includegraphics[width=.47\textwidth]{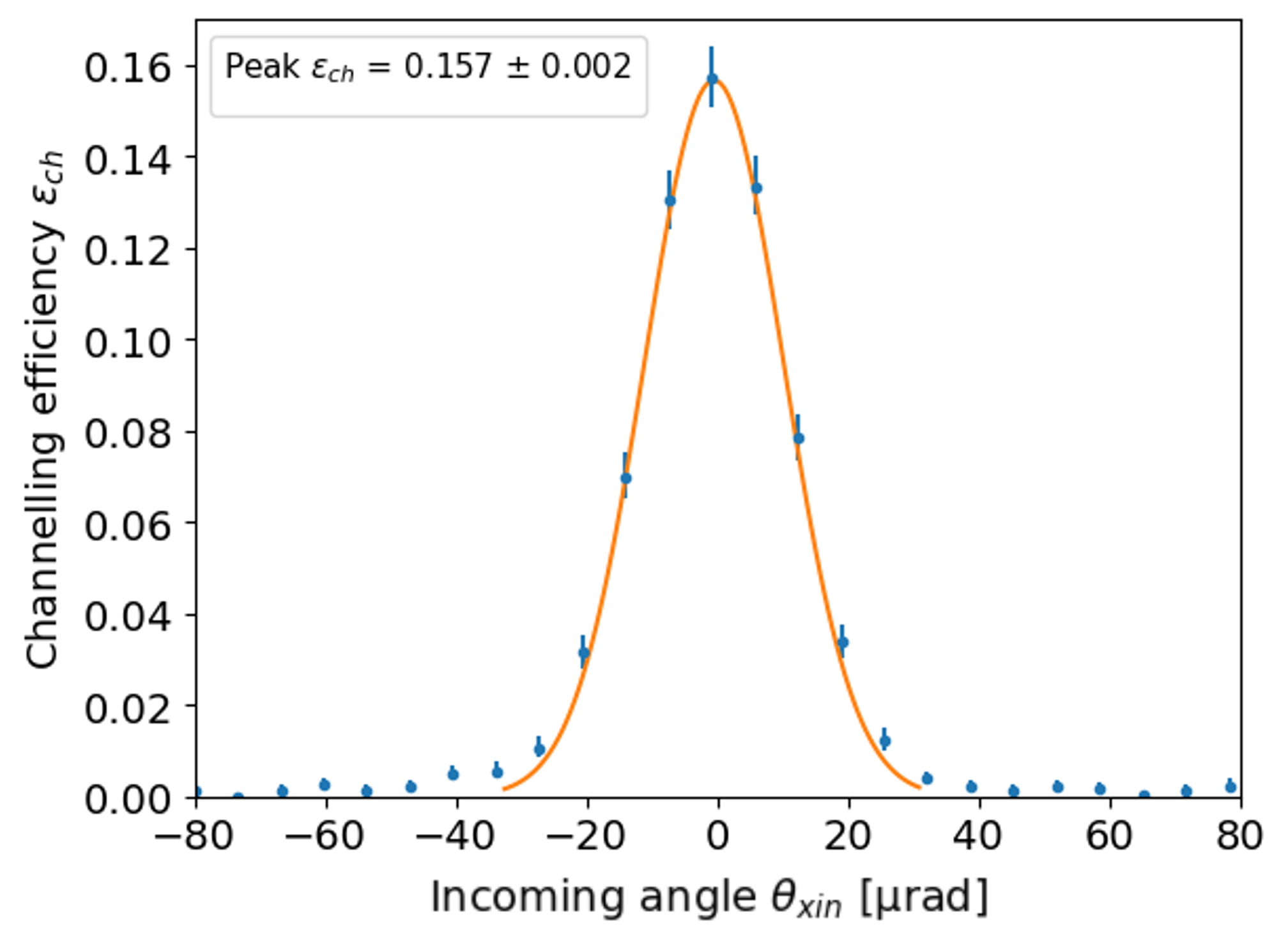}} \\
    \caption{TCCP (top) and TCCPA (bottom) crystal efficiency as a function of the incoming angle, after application of the torsion correction. The value of the peak is the efficiency of the crystal when it is perfectly aligned to the beam.}
    \label{eff_aftertorsion}
\end{figure}

For the cases of the TCCP and TCCPA crystals, the data sample is divided into sub-samples, as a function of the particle impact position on the crystal surface. For each sub-sample, the channelling efficiency is derived by calculating the ratio of the particles under the channelled peak and the total number of particles considered in the selected angular range. In this case the two peaks (VR and channelled) are very well separated and the dechannelled background is negligible, because it is spread over a much broader angular range. The angular range for which particles are considered, is defined, as for the TCCS, as all particles within $-1/2\ \theta_L< \theta_\text{in}<1/2\ \theta_L$, see Figure~\ref{eff_aftertorsion}. This allows identifying the orientation with the maximum channelling efficiency, corresponding to the orientation of the channelling planes of the crystal. 

\begin{figure}[!ht]
  \includegraphics[width=.49\textwidth]{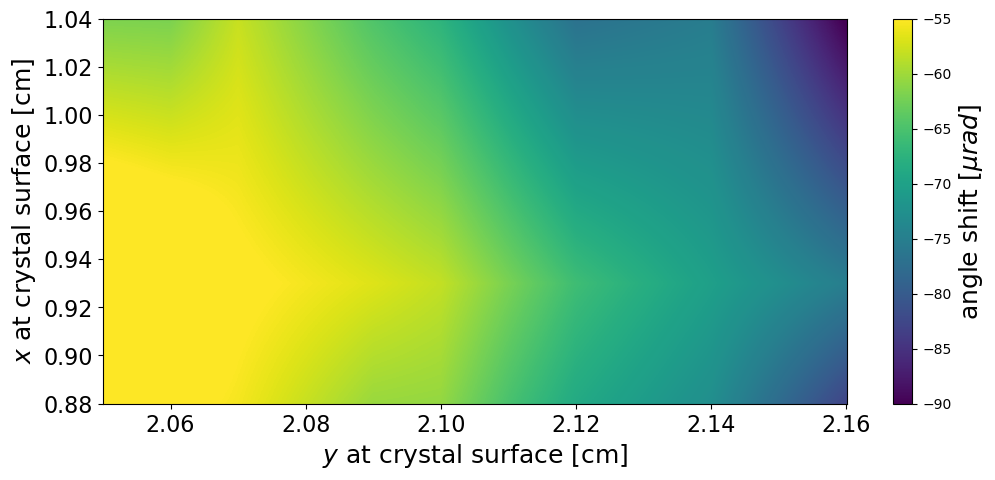} \\
  \includegraphics[width=.49\textwidth]{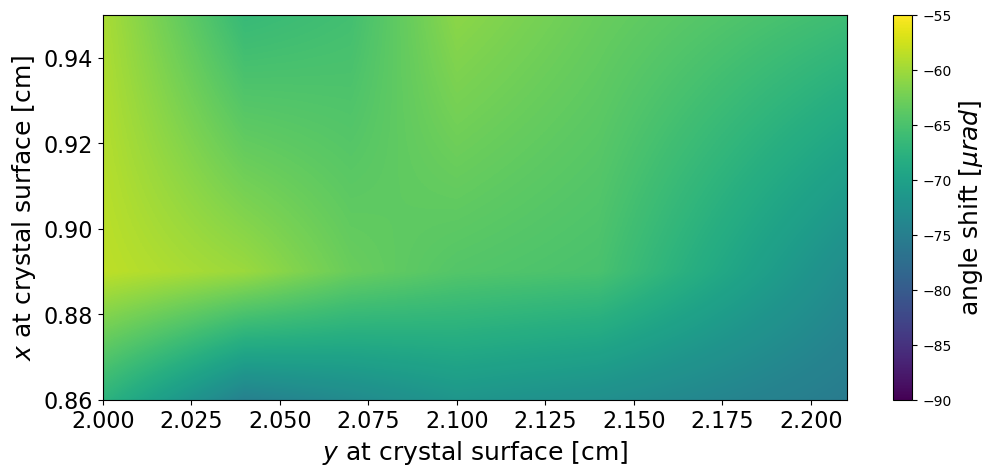} \\
\caption{Torsion map of the entrance surface of the TCCP (top) and TCCPA (bottom) crystals. The angle shift quantifies the orientation offset of the crystallographic planes along the surface with respect to the theoretical one. Despite of the smaller bending angle, the TCCP crystal torsion is less homogeneous than that of the TCCPA.}
\label{fig:torsion}       
\end{figure}

The 2D torsion maps obtained for both crystals are shown in Figure~\ref{fig:torsion}. These values are then fitted using a linear interpolator along the \textit{x} and \textit{y} in the entire crystal surface to extract the continuous distribution of the map.
The average values for the torsion on the \textit{y} direction of the two crystals are reported in Table \ref{tab:3crystals_meas}. The torsion along the \textit{x} direction is negligible, therefore is not reported. The torsion of the TCCPA crystal is lower by a factor of five with respect to the TCCP, which is an excellent result and shows the anodic-bonding technique is promising for large-angle bent crystals.

The final efficiency curve is computed after applying an angular shift to the incoming direction of the particles, accordingly to the value of the torsion map at their
impact position on the crystal surface. This way we obtain a global efficiency curve for all the particles in our dataset. The two curves for the TCCP and TCCPA crystals are shown in Figure~\ref{eff_aftertorsion}. The peak efficiency values computed for all three crystals are reported in Table~\ref{tab:3crystals_meas}. For the case of the TCCPA crystal, the procedure to remove the torsion effect is more effective than in the case of the TCCP. This is visible in the peak of the efficiency curve in Figure~\ref{eff_aftertorsion}, which is perfectly centred at zero. This likely results from second-order torsion effects in TCCP and a non-uniform bending radius along the \textit{z} direction, which was not observed in the TCCPA crystal.

\begin{figure}[h]
  \centering
 \includegraphics[width=.47\textwidth]{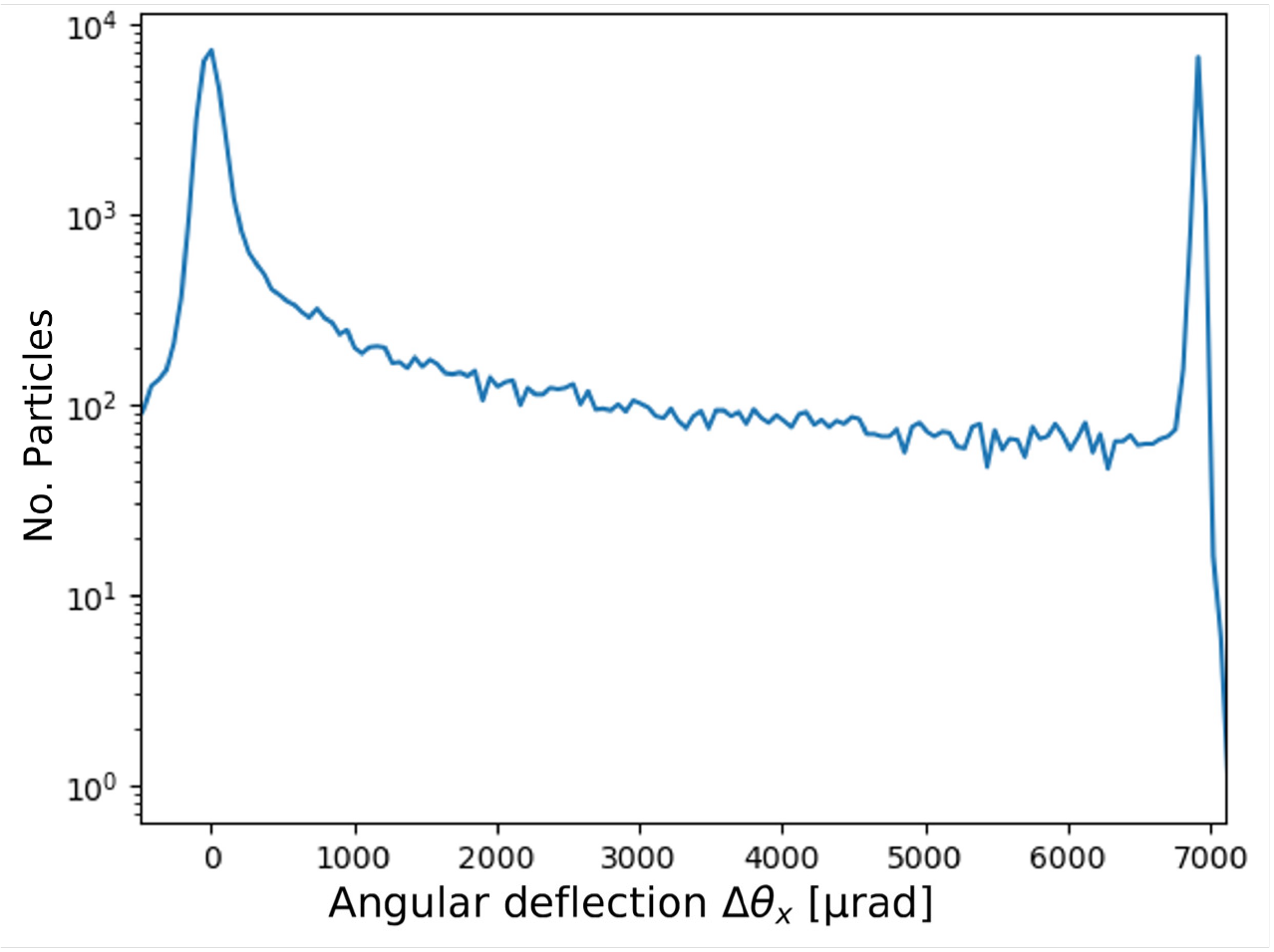}\\
\caption{Distribution of the angular deflection of particles which impact the TCCP crystal within half the Lindhard angle $(12.9/2)\SI{}{\micro\radian}$.}
\label{fig:deflection_TCCP}
\end{figure}

The bend angle $\theta_b$ of all the crystals is computed by fitting the channelling peak and extracting the mean of the peak. In this case, the background due to the dechannelled particles is negligible, as is shown in Figure~\ref{fig:deflection_TCCP}. The value obtained agrees with the technical specifications in Table~\ref{tab:3crystals}.

The measured peak efficiency of the TCCP crystal yields 15.8\%. This is discussed in detail, and compared to simulations, in the next section. 

The measured efficiency of the TCCPA crystal is comparable to that of the TCCP. This is a remarkable result, considering that the bending angle is almost twice as high as for the TCCP. 

With a measured efficiency of \SI{61.9 \pm 0.5}{\percent}, the TCCS crystal performs similarly to crystals used for beam collimation~\cite{dandrea_characterization_2023}. 

\begin{table}
\caption{Measured properties of the three crystals. Torsion was not measured for the TCCS due to the negligible effect.}
\resizebox{1.0\columnwidth}{!}{%
\label{tab:3crystals_meas}
\begin{tabular}{lccc}
\hline\noalign{\smallskip}
Crystal & $\epsilon_\text{ch}$ [$\%$] & $\theta_{b}$ {[}\SI{}{\micro\radian}{]} & $\tau_{y}$ {[}\SI{}{\micro\radian/\milli\meter}{]}  \\
\noalign{\smallskip}\hline\noalign{\smallskip}
TCCS& \SI{61.9(0.5)}{} & \SI{48.2(0.1)}{} & -\\ 
TCCP& $15.8\pm0.1$ & $6921.3\pm 0.6$ & $24.4 \pm 0.2$ \\
TCCPA & $15.7\pm0.2$ & $13295\pm 2$& $4.6\pm 0.2$ \\
\noalign{\smallskip}\hline
\end{tabular}}
\end{table}

\section{Simulations of the hadron beam experiment}
\label{sec:simulation}
To predict the expected channelling efficiency in the bent crystals, simulations of the hadron beam test were carried out by using the crystal routine~\cite{Redaelli:2018zjm,mirarchi_crystal_2018,dandrea_release_2021} integrated into the particle tracking code Xsuite~\cite{iadarola_xsuite_2023}. Simulations are important to draw conclusions on the quality of the crystals as measured in hadron-beam validation, as poor channelling efficiency values can indicate issues with the crystal quality. Xsuite is a simulation package recently developed at CERN, designed to replace various tracking codes, including SixTrack, the previously used code for crystal collimation simulations~\cite{Redaelli:2018zjm}. Xsuite employs a modular format with just-in-time compilation of \texttt{C} into \texttt{Python}, enabling more streamlined development and faster simulations. The crystal simulation routine in Xsuite is an updated implementation of previous models, incorporating multiple interactions along the length of the crystal~\cite{demetriadou_tools_2023,van_der_veken_recent_2024}.
%
%
%
Comparative simulations of the single-pass experiment performed in the H8 beamline, using both SixTrack and Xsuite, are reported in~\cite{dewhurst_calculating_2024}. In this paper, we focus on Xsuite simulations, as the multiple-interaction feature is expected to benefit the study of longer crystals such as the TCCP and TCCPA. 

\subsection{Simulation setup and analysis approach}

To simulate the particle channelling experiment, a beam distribution of \SI{e7}{} particles is initialised at the crystal entrance with Gaussian transverse profiles of $\sigma_{x,y} = \SI{2}{\milli\meter}$ and $\sigma_{x',y'} = \SI{60}{\micro\radian}$ based on the hadron beam distributions (\textit{reference} dataset) measured in the H8 beamline. Xsuite is incapable of simulating pion beams, so only protons are simulated. They have the same unit positive charge and interact with the crystalline potential in a similar way. 

The simulation assumes perfect crystals with uniform bending, and no torsion, miscut, or amorphous layer at the crystal entrance/exit surfaces. Therefore, it provides an upper bound for channelling efficiency expected from experimental measurements. 

After the simulated beam profiles are recorded, the expected angular resolution of the detector system is applied to the incoming and outgoing angles as a post-processing step, as done in previous similar analyses~\cite{mirarchi_crystal_2015-1}. The expected resolution of \SI{4.5}{\micro\radian} for incoming angles and \SI{7.2}{\micro\radian} for outgoing angles is an estimate, based on simulations, with the main contribution arising from multiple scattering in the air surrounding the crystal and each detector. This resolution is applied as a random noise signal by convoluting the incoming and outgoing trajectories from the simulations with a Gaussian of standard deviation $\sigma$ equal to the resolution.

\begin{figure}
  \includegraphics[width=.49\textwidth]{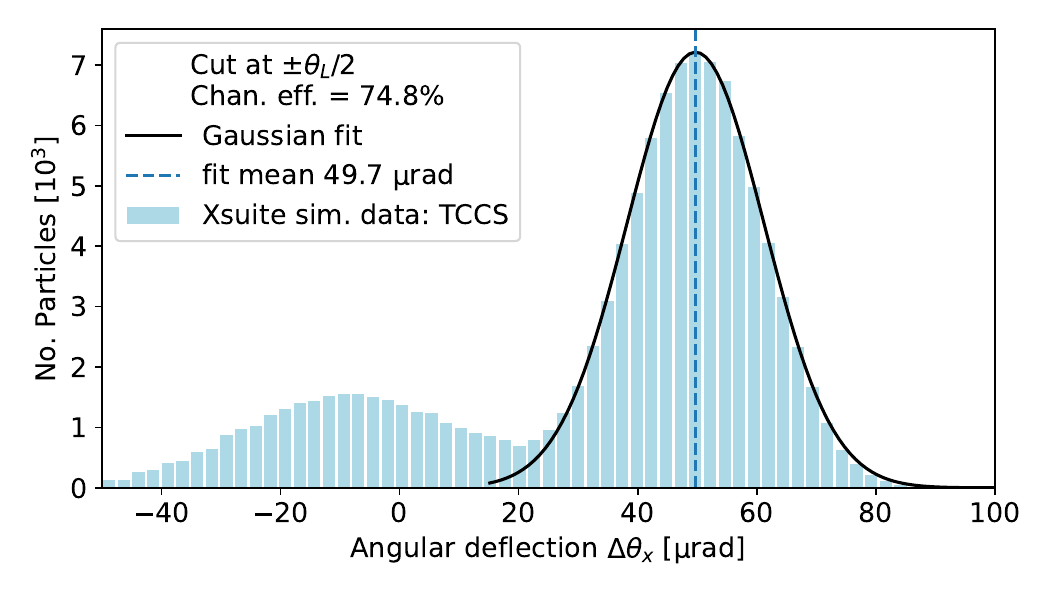}
    \caption{Xsuite simulation result, showing the angular deflection of particles which impact the TCCS crystal within half the Lindhard angle. The bin width is \SI{2.5}{\micro\radian}, consistent with Figure~\ref{fig:TCCS_ch_eff}. A Gaussian fit performed on the channelled peak data has a mean of \SI{49.8}{\micro\radian} and standard deviation of \SI{12.2}{\micro\radian}. The area enclosed by the full Gaussian curve contains \SI{76.9}{\%} of the deflected particles.}
\label{fig:sim_TCCS_ch_eff}
\end{figure}

The channelling efficiency is derived based on the histograms of the angular deflection. A Gaussian fit is performed on the right side of the channelled peak, in full analogy to the analysis of the measured data. The simulated distribution of angular deflection in the TCCS is shown in Figure~\ref{fig:sim_TCCS_ch_eff}. It is the equivalent to the data shown in Figure~\ref{fig:TCCS_ch_eff}, used to estimate the measured channelling efficiency. 

A small difference in channelling efficiency is expected due to the different masses of protons and $\pi^+$-hadrons. The dechannelling length~\cite{biryukov_crystal_1997} at a momentum of \SI{180}{\giga\electronvolt} is 12\% lower for $\pi^+$-hadrons than for protons. The associated reduction in channelling efficiency with the mixed H8 beams compared to proton beams can then be calculated to be 3\% for the TCCP and TCCPA, and less than 1\% for the TCCS. The results reported in the following sections will include a corresponding correction to account for the fact that the simulations were performed with protons and not with the mixed hadron H8 beams.

\subsection{Simulation results}

\subsubsection{TCCS Crystal}
A summary of the simulations and a comparison with the experimental results is provided in Table~\ref{tab:sim_efficiency}. For the TCCS crystal, the simulated efficiency is found to be 74.8\%, whereas the measured efficiency is 61.9$\pm0.5$\%. The simulation thus overestimates the channelling efficiency compared to the experimental result. Several factors may contribute to the observed difference between simulated and measured efficiencies. One possible explanation is an underestimation of the detector resolution in the simulation. Sensitivity studies indicate that if the angular resolution of the detectors are assumed to be \SI{9}{\micro\radian} instead of the originally simulated values, the simulation result would align more closely with the measured efficiency. 

Past measurements~\cite{matheson_presentation_22} of comparable crystals have reported a broad range of measured efficiencies, between 52\% and 69\%. The measured efficiency for the TCCS crystal in this study remains well within this range although it does not reach the performance of the crystals used for beam collimation. The discrepancy between the simulated and measured efficiency could therefore be attributed to inherent differences in crystal quality, minor surface imperfections, or variations in experimental conditions. The crystal is nevertheless deemed suitable for installation into the LHC.


\subsubsection{TCCP and TCCPA Crystals}

The performance of the TCCP and TCCPA crystals shows a larger discrepancy between simulation and measurement compared to the TCCS crystal. For the TCCP, the simulated channelling efficiency is 36.6\%, whereas the measured efficiency is significantly lower at 15.8\%. Similarly, for the TCCPA, the simulated efficiency is 29.2\%, compared to the measured value of 15.9\%. This discrepancy is notably larger than for the TCCS crystal and cannot be attributed to differences in detector resolution alone. For the larger deflection angles, compared to the TCCs, the impact of the resolution, being in the \SI{}{\micro\radian} range, is expected to be small.

To investigate this discrepancy, various sensitivity studies are performed within the simulation framework. Different initial beam distributions are tested, and the sensitivity to detector resolution is re-evaluated. These studies showed only a small dependence on resolution, with a 0.25 percentage-point change per \SI{}{\micro\radian} of resolution degradation. Additionally, variations in beam parameters (e.g., beam size) resulted in notable efficiency changes only when applying drastic modifications that would be unrealistic to assume under experimental conditions. It should also be noted that there is currently limited experience with Xcoll in the context of this particular type of long crystal, as benchmarking so far has primarily been performed against the performance of short crystals, such as the TCCS.




\begin{table}
\begin{tabular}{lccc}
%
%
\hline\noalign{\smallskip}
Crystal & $\theta_{bend}$ [\SI{}{\micro\radian}] & $\epsilon_\text{sim}$ [\%] & $\epsilon_\text{exp}$ [\%]  \\
\hline
TCCS& \SI{50}{} & \SI{74.8}{}  & $61.9\pm 0.5$ \\ 
TCCP& \SI{7000}{} & \SI{36.6}{}  & $15.8\pm 0.1$\\
TCCPA & \SI{13300}{} & \SI{29.2}{}  & $15.9\pm 0.2$\\
\noalign{\smallskip}\hline
\end{tabular}
\caption{Simulated channelling efficiency, $\epsilon_\text{sim}$, for each crystal using Xsuite compared with the measured channelling efficiency, $\epsilon_\text{exp}$. For comparison, two different values of the interplanar crystal potential are used in simulations. The simulated efficiencies reported here are, for the long crystals, reduced by 3\% to take into account the reduced efficiency expected for the mixed \SI{180}{\giga\electronvolt} $\pi^+$-hadron and proton beams.}
\label{tab:sim_efficiency}
\end{table}

A potential loss of efficiency due to variations in the bending radii, which is generally expected to reduce efficiency, cannot explain the discrepancy, as it is consistently observed in both the TCCP and TCCPA, despite their very different bending radius variations along their length.

It should be noted that the simulation tool incorporates fundamental physical concepts of a Monte Carlo framework, including the commonly used concept of dechannelling length, and has been benchmarked against crystals of several millimeters in length. However, the TCCP and TCCPA crystals are an order of magnitude longer than those previously studied. One possible explanation for the discrepancy is that dechannelling due to small non-conformities may occur more frequently in longer crystals than in the TCCS, making the overestimation of channelling efficiency in a simulation assuming a perfect crystal more pronounced. Since the Xsuite code does not explicitly model local imperfections that could contribute to dechannelling along the crystal length, the simulation may overestimate the efficiency of longer crystals more significantly than shorter ones.

Further investigations and benchmarking against other Monte Carlo tools will be performed to identify potential sources of the observed differences. The findings further underline the scientific interest in understanding the crystal channelling behaviour in the LHC, as this could deliver crucial input to further improve exisiting simulation tools.

\section{Conclusions}
\label{sec:concl} 
The TWOCRYST experiment aims to demonstrate the feasibility of using bent crystals for double channelling at the LHC in the TeV energy range. In preparation for its implementation, three silicon bent crystals - TCCS, TCCP, and TCCPA - were produced and extensively tested at the CERN SPS extraction lines with \SI{180}{\giga\electronvolt} hadrons. The crystals were characterised using both high-resolution X-ray diffraction and hadron beam measurements, providing detailed insight into their bending angles, torsion profiles, and channelling efficiencies. The experimental results were compared with Monte Carlo simulations performed using Xsuite.

The TCCS crystal exhibited a measured channelling efficiency of \SI{61.9 \pm 0.5}{\percent}, in good agreement with expectations based on previous measurements of crystals with the same specifications used for LHC beam collimation. The TCCP and TCCPA crystals, which feature longer lengths and significantly larger bending angles, showed a measured channelling efficiency of 15.8\% and 15.9\%, notably lower than the corresponding simulations, which predicted 36.6\% and 29.2\%. Unlike the TCCS, this discrepancy cannot be attributed to detector resolution effects alone. Further studies, including benchmarking against other Monte Carlo tools and incorporating more detailed crystal deformation models, will be conducted to refine the predictions. The TCCPA crystal is not planned to be used in TWOCRYST. However, this technology shows a very strong potential for particle accelerator applications because this crystal was demonstrated to provide smaller torsion and larger channelign efficiencies. 

Despite the discrepancies in absolute channelling efficiency values, both TCCS and TCCP were deemed suitable for installation in the LHC for the TWOCRYST experiment. The TCCS meets expectations based on past experience with collimation crystals, while the TCCP, despite its lower measured efficiency, still demonstrates feasibility for spin-precession beam experiments planned by TWOCRYST. These results in fact strengthen further the need to assess the performance at higher energies, as close as possible to those of a final LHC experiment. The results obtained in this study will complemented by measurement in the multi-TeV energy range and be used as input for future high-energy applications of bent crystals, including for the ALADDIN experiment, and help refine simulation models for long crystals in high-energy accelerators.



%

\begin{acknowledgements}
The authors acknowledge the CERN Experimental Areas group for their support for the measurements in the CERN North Area.
This work received support from the ERC SELDOM Consolidator Grant G.A. n. 771642.

\end{acknowledgements}

\bibliographystyle{spphys}       

\bibliography{TWOCRYST_H8}


\end{document}